\documentclass[final,5p,times,twocolumn,preprintnumbers,superscriptaddress,numbers,square]{elsarticle}
\usepackage[T1]{fontenc}   
\usepackage{amssymb,amsmath,amsfonts,accents}
\usepackage{xcolor}
\usepackage{soul}
\usepackage[inline]{enumitem}
\usepackage{graphicx}
\usepackage[breaklinks]{hyperref}
\usepackage{orcidlink}
\hypersetup{
 colorlinks,
 linkcolor={red},
 citecolor={blue},
 urlcolor={blue}
}
\usepackage{xfrac}
\usepackage{mathpazo}
\biboptions{sort&compress}
\journal{Physics Letters B}
\begin{document}
\begin{frontmatter}

\title{Regularity of core coupled high-spin structure in neutron-rich $^{196\mbox{-}202}$Au isotopes approaching N $=126$}

\author[1,2,3]{Y.~Cho~\orcidlink{0009-0000-1767-996X}~\fnref{curraddr}}
\affiliation[1]{addressline={Department of Physics and Astronomy, Seoul National University, Seoul 08826, Republic of Korea}}
\affiliation[2]{addressline={Institute for Nuclear and Particle Astrophysics, Seoul National University, Seoul 08826, Republic of Korea}}
\affiliation[3]{addressline={Center for Exotic Nuclear Studies, Institute for Basic Science, Daejeon 34126, Republic of Korea}}
\fntext[curraddr]{Present address: Physics Division, Argonne National Laboratory, Argonne, Illinois 60439, USA}
\author[3]{Y.~H.~Kim~\orcidlink{0000-0002-7221-058X}~\corref{c2}}
\cortext[c2]{Corresponding author}
\ead{yunghee.kim@ibs.re.kr}
\author[4]{A.~Navin~\orcidlink{0000-0003-2502-5864}}
\affiliation[4]{addressline={Grand Accélérateur National d'Ions Lourds, CEA/DRF-CNRS/IN2P3, Boulevard Henri Becquerel, F-14076 Caen, France}}
\author[4]{M.~Rejmund\orcidlink{0009-0009-8626-8756}}
\author[4]{A.~Lemasson\orcidlink{0000-0002-9434-8520}}
\author[4]{D.~Ramos~\orcidlink{0000-0003-1822-6537}}
\author[4]{E.~Clément~\orcidlink{0000-0003-1887-717X}}
\author[5]{J.~Dudouet~\orcidlink{0000-0002-9018-6763}}
\affiliation[5]{addressline={Université Lyon, Université Claude Bernard Lyon 1, CNRS/IN2P3, IP2I Lyon, UMR 5822, F-69622 Villeurbanne, France}}
\author[4]{P.~Van~Isacker~\orcidlink{0000-0003-3938-9220}}
\author[6]{C.~Yuan~\orcidlink{0000-0002-3495-3614}}
\affiliation[6]{addressline={Sino-French Institute of Nuclear Engineering and Technology, Sun Yat-Sen University, Zhuhai, Guangdong 519082, China}}
\author[6]{M.~Liu~\orcidlink{0000-0001-8028-1487}\fnref{curraddr2}}
\fntext[curraddr2]{Present address: Center for Exotic Nuclear Studies, Institute for Basic Science, Daejeon 34126, Republic of Korea}
\author[7]{A.~N.~Andreyev~\orcidlink{0000-0003-2828-0262}}
\affiliation[7]{addressline={School of Physics, Engineering and Technology, University of York, Heslington, York, North Yorkshire YO10 5DD, United Kingdom}}
\author[1,2,3]{Y.~Son~\orcidlink{0009-0007-7117-4357}}
\author[1,2]{S.~Choi\orcidlink{0000-0002-9448-969X}}
\author[4]{D.~Ackermann~\orcidlink{0000-0001-6284-1516}}
\author[8]{G.~de Angelis~\orcidlink{0000-0002-0506-3815}}
\affiliation[8]{addressline={INFN Laboratori Nazionali di Legnaro, IT-35020 Legnaro, Italy}}
\author[3]{S.~Bae~\orcidlink{0000-0003-0420-2163}}
\author[9]{R.~Banik~\orcidlink{0000-0001-6969-0458}}
\affiliation[9]{addressline={Institute of Engineering \& Management, Kolkata 700091, India}}
\author[10]{S.~Bhattacharya~\orcidlink{0000-0002-7198-660X}}
\affiliation[10]{addressline={Variable Energy Cyclotron Centre, 1/AF, Bidhannagar, Kolkata, West Bengal 700064, India}}
\author[10]{S.~Bhattacharyya~\orcidlink{0000-0002-1712-5825}}
\author[11]{K.~Chae~\orcidlink{0000-0003-1819-079X}}
\affiliation[11]{addressline={Department of Physics, Sungkyunkwan University, Suwon 16419, Korea}}
\author[12]{F.~Didierjean~\orcidlink{0009-0002-4950-3162}}
\affiliation[12]{addressline={University of Strasbourg, CNRS/IPHC UMR7178, F-67000 Strasbourg, France}}
\author[4]{C.~ Fougères~\orcidlink{0000-0002-1236-4739}}
\author[4]{G.~de France~\orcidlink{0000-0002-7439-1759}}
\author[4]{G.~Fremont}
\author[4]{J.~Goupil}
\author[3,11]{G.~M.~Gu~\orcidlink{0009-0007-5112-8505}}
\author[3,13]{J.~Ha~\orcidlink{0000-0003-4843-8573}}
\affiliation[13]{addressline={University of Padua, Via 8 Febbraio, 1848, 2, 35122 Padua, Italy}} 
\author[3]{K.~I.~Hahn~\orcidlink{0000-0002-9250-6484}}
\author[14]{Y.~Hirayama~\orcidlink{0000-0002-9485-1177}}
\affiliation[14]{addressline={Wako Nuclear Science Center, IPNS, High Energy Accelerator Research Organization (KEK), Wako, Saitama 351-0198, Japan}}
\author[14]{S.~C.~Jeong}
\author[15]{D.~T.~Kattikat~Melcom~\orcidlink{0009-0001-0427-5584}}
\affiliation[15]{addressline={LP2I Bordeaux, UMR5797, Université de Bordeaux, CNRS, F-33170, Gradignan, France}}
\author[11]{C.~H.~Kim~\orcidlink{0000-0002-4614-4793}}
\author[3,11]{M.~J.~Kim~\orcidlink{0000-0002-4372-5592}}
\author[3]{S.~H.~Kim~\orcidlink{0009-0008-6013-5982}}
\author[16]{W.~Korten~\orcidlink{0000-0002-3940-0816}}
\affiliation[16]{addressline={Université Paris-Saclay, IRFU, CEA, Gif-sur-Yvette, F-91191, France}}
\author[4,15]{P.~Marini~\orcidlink{0000-0001-9537-8126}}
\author[14]{H.~Miyatake~\orcidlink{0000-0002-6888-8507}}
\author[17]{M.~Mukai~\orcidlink{0000-0002-7146-2961}~\fnref{curraddr3}}
\fntext[curraddr3]{Present address: Wako Nuclear Science Center, IPNS, High Energy Accelerator Research Organization (KEK), Wako, Saitama 351-0198, Japan}
\affiliation[17]{RIKEN Nishina Center, 2-1 Hirosawa, Wako, Saitama 351-0198, Japan} 
\author[3]{A.~Mukherjee~\orcidlink{0009-0009-3590-3668}}
\author[10]{G.~Mukherjee~\orcidlink{0000-0003-0457-4012}}
\author[14]{T.~Niwase~\orcidlink{0000-0002-0054-2278}}
\author[18]{T.~Otsuka~\orcidlink{0000-0002-1593-5322}}
\affiliation[18]{addressline={Department of Physics, University of Tokyo, Hongo, Bunkyo-ku, Tokyo 113-0033, Japan}}
\author[3]{J.~Park~\orcidlink{0000-0003-0866-7523}~\fnref{curraddr4}}
\fntext[curraddr4]{Present address: Department of Physics, Hope College, Michigan 49423, United States}
\author[19]{R.~M.~Pérez-Vidal~\orcidlink{0000-0002-4075-4152}}
\affiliation[19]{addressline={Instituto de Física Corpuscular, CSIC-Universidad de Valencia, Valencia E-46980, Spain}} 
\author[20]{F.~Recchia~\orcidlink{0000-0002-8428-0112}}
\affiliation[20]{addressline={Dipartimento di Fisica e Astronomia, Universitá Degli Studi di Padova, Padova, Italy}}
\author[21]{K.~Rezynkina~\orcidlink{0000-0002-3017-6440}}
\affiliation[21]{addressline={Istituto Nazionale di Fisica Nucleare, Sezione di Padua, I-35131 Padua, Italy}}
\author[17]{P.~Schury~\orcidlink{0000-0002-7094-6761}~\fnref{curraddr3}}
\author[4]{J. C.~Thomas~\orcidlink{0000-0001-6618-9772}}
\author[15]{I.~Tsekhanovich~\orcidlink{0000-0002-8188-544X}}
\author[4]{A.~Utepov~\orcidlink{0000-0002-1015-7285}}
\author[14]{Y.~X.~Watanabe~\orcidlink{0000-0003-2425-7188}}
\date{\today}

\begin{abstract}
Excited states built upon long-lived high-spin isomers in hard-to-reach neutron-rich Au ($Z=79$) isotopes
were populated using multi-nucleon transfer reactions between $^{136}$Xe and $^{198}$Pt at 7 MeV/u. 
These states in $^{196,\,198-202}$Au were identified for the first time using the powerful combination of the VAMOS++ spectrometer, the CATLIFE detection system, and the AGATA $\gamma$-ray tracking array.
Their measured energies exhibit remarkable regularity as a function of neutron number and are seen to be inherited from the energies of yrast-band members in the corresponding Hg ($Z=80$) isotope.
Large-scale shell-model calculations reproduce the observed regularity and show that these states arise from the unique-parity orbitals $\pi0h_{11/2}$ and $\nu0i_{13/2}$ coupled to the Hg core. 
This regularity is due to the dominant proton configurations of the Hg and Au isotopes, where the level energies are almost independent of different neutron-orbital occupancies.
The calculated Au wave functions show significant higher-spin components of the corresponding Hg core, unlike what is expected in the conventional interpretation in terms of the weak-coupling/decoupling limits of the particle-core coupling model.
\end{abstract}
\end{frontmatter}
In spherical nuclei, near the major neutron and proton shell closures, nuclear excitations arise from the arrangement of quantum orbitals within the nuclear potential well and the associated individual (single-particle) motion of the valence nucleons near the nuclear surface~\cite{Sorlin2008,Otsuka2020}. 
As the number of active valence protons and neutrons increases, the strong correlations among the nucleons favor a coherent collective motion, resulting in deformed nuclei~\cite{Heyde2011, Butler2016}.
Nuclei can also undergo vibrations or rotations, analogous to those of molecules. 

High-spin excited states of odd-even or odd-odd nuclei typically arise from the interplay between the odd nucleons in high-$j$ orbitals and the collective behavior of the underlying core. 
In this context, Au ($Z=79$) isotopes are well suited to explore how collective nuclear modes evolve along with high-$j$ single-particle excitations over a long isotopic chain. 
In the following, we briefly describe the currently known high-spin structure of Au isotopes beyond $^{189}$Au:
The ground state $J^\pi$ of the even-$N$ $^{191-203}$Au is $3/2^+$, which is related to the $\pi 1d_{3/2}$ orbital. 
Band structures built on the often isomeric $11/2^-$ states in even-$N$ $^{189\text{–}197}$Au~\cite{gono1977, gono1979, TJOM1974, kolscbach1985, Gueorguieva2003, Oktem2007, WANG2012, DRACOULIS2013, Wheldon2006} and on the $11^-$ states in odd-$N$ $^{190\text{–}194}$Au~\cite{neskakis1979, neskakis1982, Gueorguieva2004, oktem2012, gao2012} have been observed.
These have been interpreted as bands based on the high-$j$ unique-parity configurations $\pi 0h_{11/2}^{-1}$ and $\pi 0h_{11/2}^{-1}\nu 0i_{13/2}^{-1}$, respectively.  The interpretation using the decoupling (rotation-aligned) limit of the particle-plus-rotor model~\cite{Bucurescu2017, Chen2006, gono1977, gono1979, TJOM1974, kolscbach1985, neskakis1979, neskakis1982, Oktem2007} and in terms of non-axial oblate shapes~\cite{Gueorguieva2003, Oktem2007, WANG2012, DRACOULIS2013, Wheldon2006, Gueorguieva2004, oktem2012, gao2012} was established based on the Hg core. 
For heavier even-$N$ isotopes, only the $11/2^-$ isomeric states ($^{199}$Au~\cite{BORGGREEN1968} and $^{203}$Au~\cite{Steer2011, PODOLYAK2009}) or a few transitions above the isomers ($^{201,\,205}$Au~\cite{Steer2011, PODOLYAK2009}) are known. In odd-$N$ $^{196-200}$Au isotopes, only $12^-$ isomeric states 
have been reported in Refs.~\cite{WAPSTRA1967, HAGN1982, Cunnane1972_198Au, PAKKANEN1973, SAKAI1968, Cunnane1972_200Au}, while high-spin states above the isomers in $^{202-204}$Au have not yet been observed. 
As discussed in Ref.~\cite{Bucurescu2017}, these heavier isotopes likely lie between the weak-coupling and the decoupling limit, given their small deformations~\cite{STONE05,ENSDF,Sahoo2024}. 
In both limits, energetically favored states $(j, j+2, j+4, ...)$ are usually interpreted as spin-$j$ particle/hole coupled to the corresponding core levels $(0^+,2^+,4^+,...)$.

Systematic experimental data on high-spin structures above isomers in neutron-rich Au nuclei near the $N\sim126$, beyond the stable $^{197}$Au, remain extremely scarce~\cite{Podolyak2023}. Different types of reactions are required to populate the high-spin states of these nuclei, since standard reaction mechanisms such as fusion evaporation, multi-fragmentation, and spallation are difficult to employ. Multi-nucleon transfer reactions, in combination with state-of-the-art instrumentation~\cite{Wheldon2006, DRACOULIS2013, John14, Watanabe2015, DESAI19, Niwase23, WATANABE25, Dasso94, Zagre08, Karpov17, Li19}, provide access to high-spin states above the isomers.
The hard-to-reach high-spin states in neutron-rich isotopes Au ($Z=79$) offer a unique opportunity to explore how high-spin collective nuclear modes evolve from the mid-shell region towards the $N=126$ shell closure.

\begin{figure}[t!]
	\centering 
	\includegraphics[width=\columnwidth]{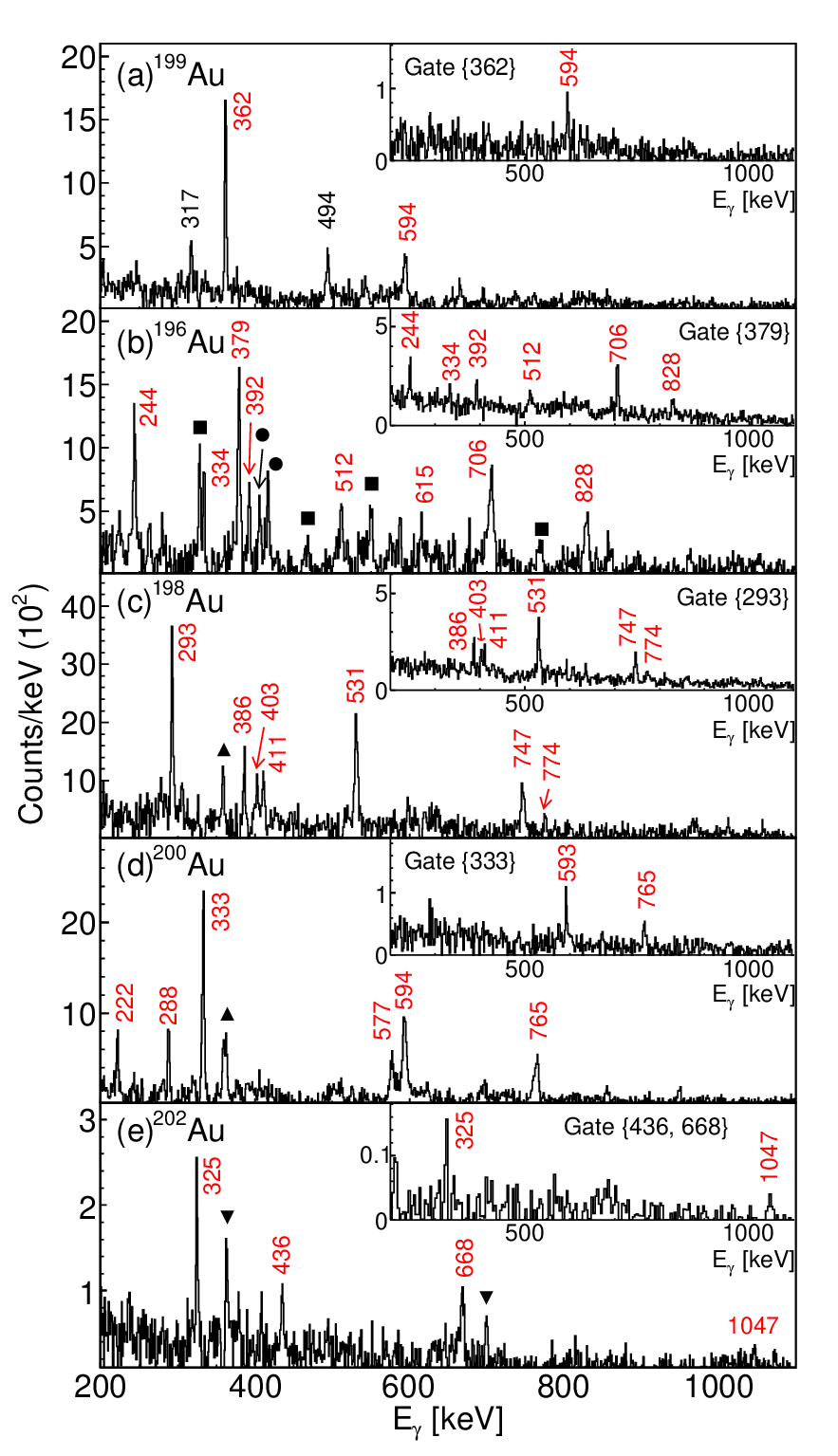}
	\caption{Prompt Doppler-corrected $\gamma$-ray spectra of Au isotopes measured in AGATA in coincidence with the complementary projectile like fragments $^{127-135}$I isotopes identified in VAMOS++: (a) $^{199}$Au, (b) $^{196}$Au, (c) $^{198}$Au, (d) $^{200}$Au, and (e) $^{202}$Au.    
    The insets show the coincidence spectra corresponding to the labeled gating transitions. The y-axis scale corresponds to counts/100 per 2 keV for (a)-(d) and per 4 keV for (e). 
    The newly identified $\gamma$-ray transitions are labeled in red. The symbols indicate the $\gamma$ rays from the neighboring Au isotopes: $^{195}$Au (circle), $^{197}$Au (square), $^{199}$Au (triangle), and $^{201}$Au (inverted triangle).
    } 
	\label{fig:GammaSpectra1}%
\end{figure}

In this Letter, we report new spectroscopic measurements of high-spin excited states in difficult-to-reach neutron-rich $^{196-202}$Au isotopes, populated via multi-nucleon transfer reactions.
These measurements show remarkable regularities in the energy of the high-spin states built on the isomer in the Au isotopes as a function of neutron number $N$.
The newly observed levels in the long chain of the Au isotopes have similar energies to those of yrast states of the corresponding Hg core. This property, which we will refer to as inheritance, is conventionally interpreted within the weak-coupling/decoupling scheme~\cite{Bucurescu2017, Oktem2007, Chen2006, gono1977, gono1979, TJOM1974, kolscbach1985, neskakis1979, neskakis1982}.
Large-scale shell-model calculations (LSSM) reproduce the observed inheritance in the high-spin states of the Au isotopes and indicate that their wave functions arise from the coupling of high-$j$ unique-parity orbitals to the Hg core.
However, substantial admixtures of higher-angular-momentum yrast states of the Hg core are found in LSSM calculations, unlike what is expected from the conventional weak-coupling/decoupling scheme.

The experiment was performed at the Grand Acc\'el\'erateur National d'Ions Lourds (GANIL), where a 7~MeV/u $^{136}$Xe beam impinging on a $1.3~\text{mg/cm}^2$ $^{198}$Pt target populated neutron-rich nuclei in the southwest of $^{208}$Pb through multi-nucleon transfer reactions~\cite{DataE806}.
The projectile-like fragments were detected using the large-acceptance VAMOS++ spectrometer~\cite{Rejmund2011, Kim2017}, positioned at $40^\circ$, near the grazing angle.
VAMOS++ enabled unambiguous isotopic identification of the projectile-like fragments and measurement of Total Kinetic Energy Loss (TKEL).
Its detection system included a Dual Position-Sensitive Multi-Wire Proportional Counters (DPS-MWPCs) at the entrance~\cite{Vandebrouck2016}, two MWPCs at the focal plane for ion tracking and timing, and a seven-fold segmented ionization chamber for atomic number ($Z$) and energy measurement. 
A machine learning method was utilized to optimize the energy resolution of the fragments detected in VAMOS++ ~\cite{CHO2023}.
The prompt $\gamma$ rays from the projectile-like and target-like fragments were detected by the Advanced GAmma Tracking Array (AGATA), comprising 36 crystals in a nominal configuration (235 mm from the target)~\cite{AKKOYUN201226, CLEMENT20171}.
The new Complementary Arm for Target LIke FragmEnts (CATLIFE)~\cite{SON2023}, which consists of an MWPC and four EXOGAM HPGe Clover detectors~\cite{Simpson2000}, was used to detect the target-like fragments and their isomeric decay transitions.

\begin{figure}[tp]
	\centering 
	\includegraphics[width=\columnwidth]{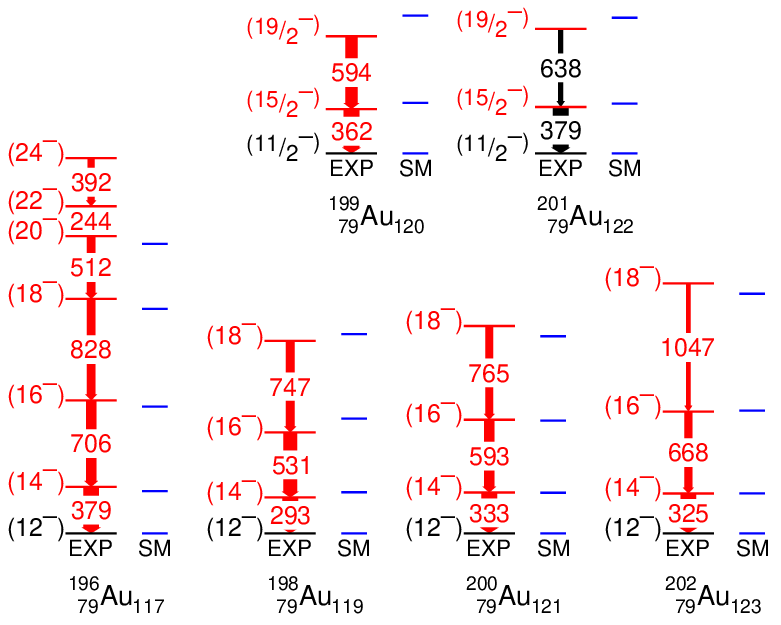}
	\caption{Partial level schemes of high-spin negative-parity states in $^{196,\,198-202}$Au isotopes obtained from the present work: The excitation energies, relative to the $11/2^-$(even-$N$) or $12^-$(odd-$N$) states, are compared to the results of the large-scale shell-model calculations indicated in blue line. The newly observed $\gamma$-ray transitions and placement of the levels are indicated in red. The thickness of the arrows indicates the intensity relative to $15/2^-\rightarrow11/2^-$ or $14^-\rightarrow12^-$ transitions. The spin parity of the shell-model calculation follows the ordering of the experimentally constructed level scheme. 
    } 
	\label{fig:LevelScheme1}%
\end{figure}

Direct particle identification of the target-like fragment is difficult due to its high $Z$, $A$ and low velocity. 
Hence, complementary projectile-like fragment information, two-body kinematics and known delayed $\gamma$-ray transitions were used to identify target-like fragment.
Prompt $\gamma$-ray transitions from target-like Au isotopes were measured by AGATA in coincidence with isotopically identified partners $^{127-135}$I, using VAMOS++.
The Doppler-shift correction for the $\gamma$ rays from the Au isotopes was carried out using 2-body kinematics for the complementary projectile-like iodine isotopes measured at VAMOS++. 
The two-dimensional $\gamma$-TKEL (Total Kinetic Energy Loss) matrices were systematically analyzed to identify transitions from the Au isotopes. 
More details are given in Ref.~\cite{CHO_PRC}. 
The prompt $\gamma$-ray spectra of the $^{196,\,198,\,199,\,200,\,201,\,202}$Au isotopes are shown in Fig.~\ref{fig:GammaSpectra1}. 
The newly observed transitions are shown in red.
Level schemes were constructed using $\gamma$-$\gamma$ coincidences and $\gamma$-ray intensities with excitation energy selections obtained from the calculated TKEL~\cite{Rejmund2007, CHO_PRC}.
The level schemes of the yrast high-spin negative-parity states are presented in Fig.~\ref{fig:LevelScheme1}. 
Extended level schemes beyond the yrast bands are given in Ref.~\cite{CHO_PRC}. 
The tentative spin assignments were guided by systematics and corroborated by LSSM calculations. 
In the case of $^{201}$Au, the ordering of two previously reported transitions~\cite{Steer2011} has been revised based on the observed intensities.

\begin{figure}[t!]
	\centering 
	\includegraphics[width=\columnwidth]{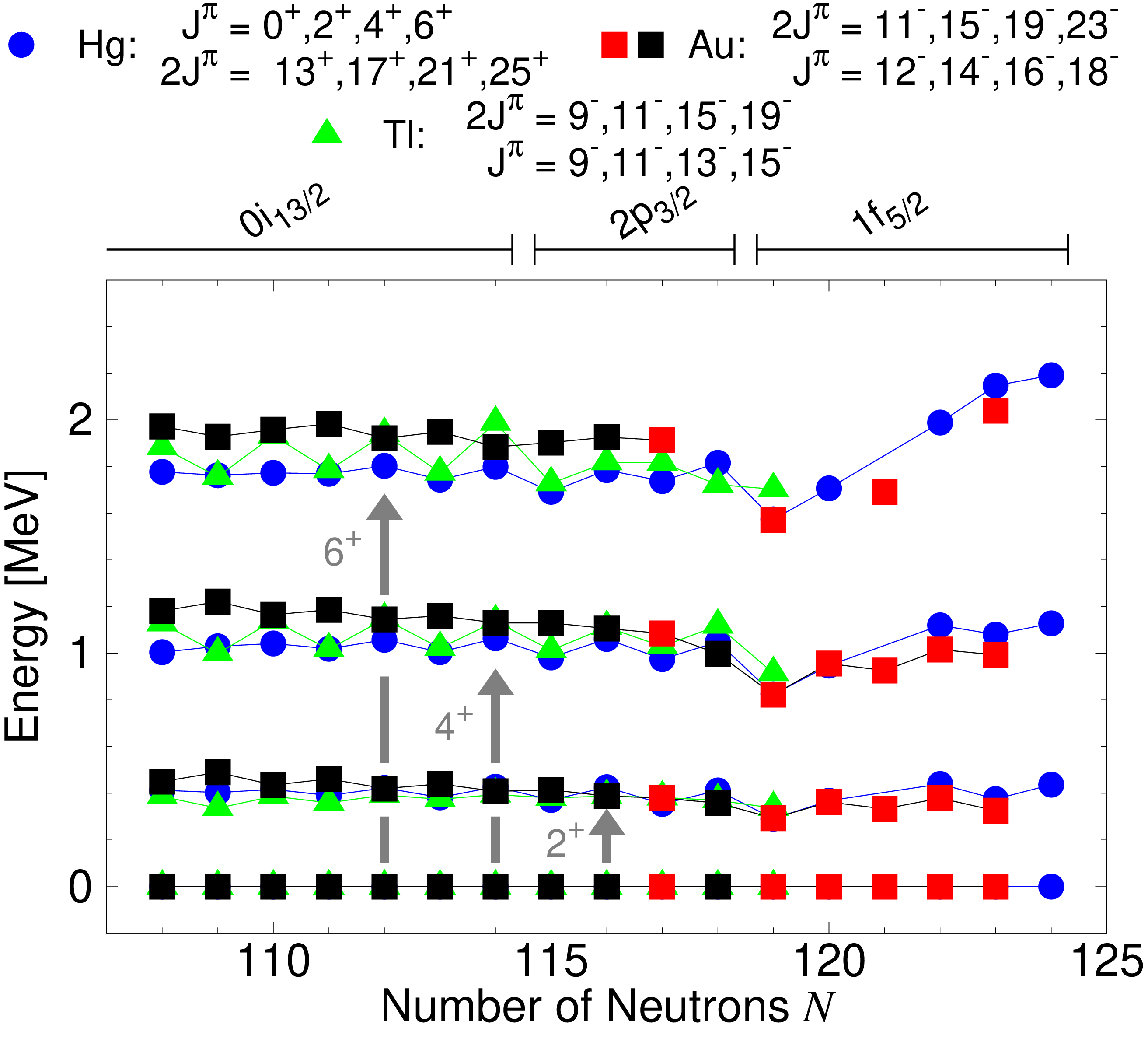}		
	\caption{
    Excitation energies of selected yrast states in Hg (blue circle), Au (red/black square) and Tl (green triangle) in $108<N<124$.
    For Au and Tl isotopes, the excitation energies are plotted relative to their respective long-lived isomeric states. 
    Arrows mark the excitation energies of the even-even Hg core states.
    The newly established energy levels in Au ($N\geq117$) are indicated in red squares; all other level energies are adopted from Ref.~\cite{ENSDF}. 
    Spin-parities of the states are indicated as $J^\pi$ ($2J^\pi$) for even-$A$ (odd-$A$) nuclei.
    The corresponding neutron orbitals involved in natural filling are also indicated.
    }
    \label{fig:Systematics}
\end{figure}

In Fig.~\ref{fig:LevelScheme1}, a pronounced regular pattern in the level energies of the high-spin states as a function of $N$ can be observed for both even-$N$ and odd-$N$ Au isotopes.
As illustrated in Fig.~\ref{fig:Systematics}, this regularity persists consistently along the long isotopic chain, spanning from the end of the shape coexistence region ($N\sim108$) to the vicinity of the $N=126$ shell closure.
In addition, it is striking that the excitation energies of the high-spin negative-parity states built on the $11/2^-$ ($12^-$) isomers in even-$N$ (odd-$N$) Au ($Z=79$) isotopes closely follow those of the corresponding excited level energies of yrast $2_1^+$, $4_1^+$, and $6_1^+$ states in Hg
($Z=80$).
This correspondence indicates a clear similarities of the level structure from the Hg core, suggesting that the observed excitations in Au isotopes are strongly influenced by the properties of even-even Hg core.
These states in Au isotopes have been interpreted to arise from configurations $\pi 0h_{11/2}^{-1}$ and $\pi 0h_{11/2}^{-1}\nu 0i_{13/2}^{-1}$, coupled to a Hg core~\cite{gono1977, gono1979, TJOM1974, kolscbach1985, Gueorguieva2003, Oktem2007, WANG2012, DRACOULIS2013, Wheldon2006, neskakis1979, neskakis1982, Gueorguieva2004, oktem2012, gao2012}.

The energies of the levels built on the $9/2^-$ ($9^-$) isomers in even-$N$ (odd-$N$) in the Tl ($Z=81$) isotopes also show systematics similar to those observed in Hg (see Fig.~\ref{fig:Systematics}).
These states in Tl isotopes have been interpreted to arise from intruder configurations~\cite{Dionisio76,Heyde1983,Wood2016}, across the $Z=82$ shell gap and correspond to the $\pi 0h_{9/2}$ (or $\pi 0h_{9/2}\nu 0i_{13/2}^{-1}$ for odd-$N$) orbital, coupled to an Hg core.
Similar structural patterns were also found in the Sn region for the $_{47}$Ag-$_{48}$Cd-$_{49}$In isotopes, highlighting the analogous roles played by the $\pi0g_{9/2}$ and $\nu0h_{11/2}$ orbitals. 

\begin{figure}[t!]
	\centering 
	\includegraphics[width=0.96\columnwidth]{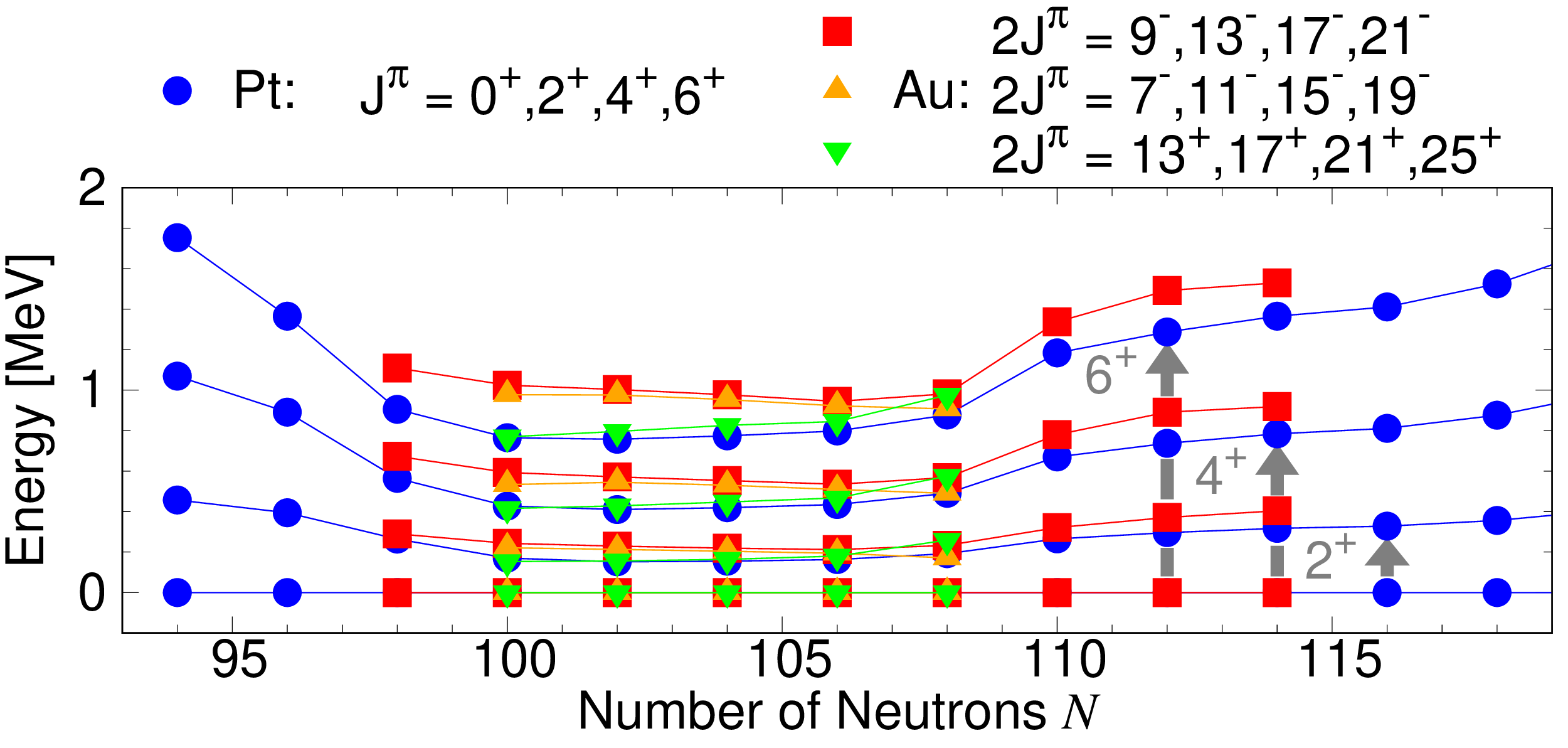}
	\caption{
     Excitation energies, relative to the bandhead, for even-$N$ Pt (circle) and for even-$N$ Au adopted from Ref.~\cite{ENSDF}.  
    Spin-parities of selected states are indicated as $J^\pi$($2J^\pi$)  for even-$A$ (odd-$A$) Pt nuclei and the $2J^\pi$ for odd-$A$ Au nuclei. The excited states of Au isotopes with configuration of the particle in one of the high-$j$ orbitals, $\pi 0h_{9/2}$ (square), $\pi 1f_{f/2}$ (triangle), or $\pi 0i_{13/2}$ (inverted triangle) with corresponding Pt core is shown.
    Arrows mark the excitation energies of the even-even Pt states.
    } 
	\label{fig:PtAu}%
\end{figure}

Analogous similarities in the level energies can also be seen between Pt and Au, as illustrated in Fig.~\ref{fig:PtAu}.
The excitation energies of the $2_1^+$, $4_1^+$, and $6_1^+$ levels in even-$N$ Pt isotopes show similar systematics with those of the states built on the $\pi 0h_{9/2}$, $\pi 1f_{7/2}$, and $\pi 0i_{13/2}$ intruder-orbital configurations in even-$N$ Au isotopes~\cite{Balogh22,Venhart17} . 
In general, Pt isotopes exhibit a higher degree of collectivity than the corresponding Hg isotone.
These intruder orbital-coupled bands in Au, observed within and near the region of shape coexistence,  also show a clear similarity with the level energies of corresponding state in  Pt ($Z=78$)~\cite{Heyde2011,WOOD1992}.

LSSM calculations were performed to gain a deeper understanding of the observed energies from the yrast states of the corresponding Hg isotope in the excited states of the Au isotopes. 
The model space encompasses the proton orbitals $\pi$($0g_{7/2}$, $1d_{5/2}$, $0h_{11/2}$, $1d_{3/2}$, and $2s_{1/2}$)
and neutron orbitals $\nu$($0h_{9/2}$, $1f_{7/2}$, $0i_{13/2}$, $2p_{3/2}$, $1f_{5/2}$, and $2p_{1/2}$).
The modified Kuo-Herling Hole Effective (KHHE) interaction~\cite{Warburton1991a,Yuan2022} was utilized with the full model space calculations until $^{198}$Au ($N=119$). 
For lighter isotopes, the monopole-based truncation with a $10$ MeV cutoff was adopted as described in Ref.~\cite{Sahoo2024, Chen2023}.
The KSHELL code~\cite{shimizu2013} was used for the computations. 
The calculated excitation energies in $^{196-202}$Au are compared to the experimental results in Fig.~\ref{fig:LevelScheme1} and show a good agreement.

\begin{figure}[t!]
\centering 
	\includegraphics[width=0.9\columnwidth]{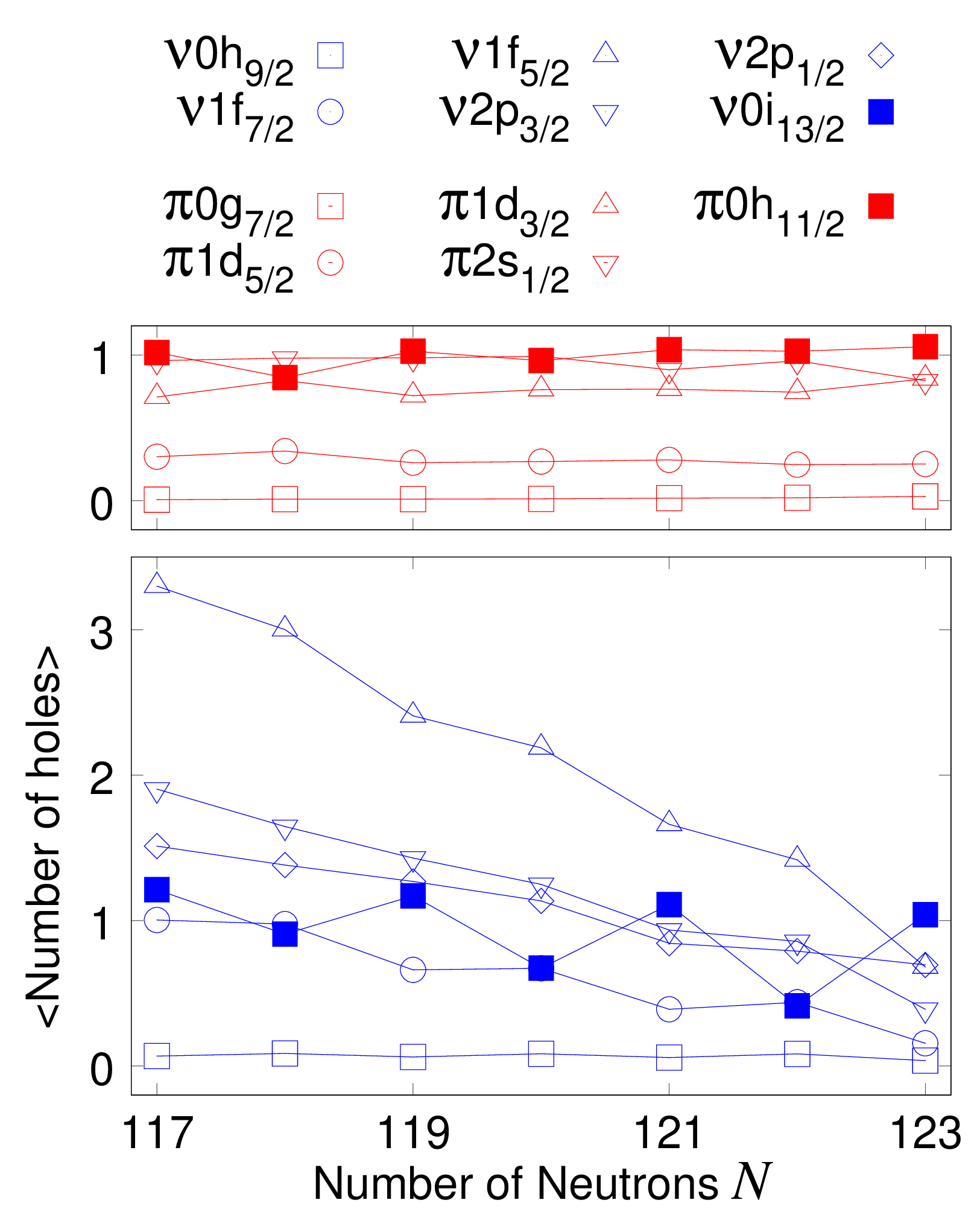}
    \caption{
    The average number of holes obtained in a monopole based truncation LSSM calculation
    for various proton (top) and neutron orbitals (bottom)
    as a function of the number $N$ of neutrons 
    for $11/2^-$ and $12^-$ states in even-$N$ and odd-$N$ isotopes of Au, respectively.
    The proton hole occupancy shows almost constant occupancy.
    In the neutron case, $\nu0i_{13/2}$ (filled blue square) unique parity orbitals presents constant number of holes over the different neutron numbers 
    and other orbitals decrease smoothly as neutron number increases.
    } 
\label{fig:Occ}
\end{figure}

The LSSM-calculated wave functions of the Au isotopes exhibit significant fragmentation among various proton and neutron configurations, reflecting the development of a substantial degree of collectivity. 
Despite this fragmentation, the proton and neutron hole occupancies for the states built on the $11/2^-$ ($12^-$) isomers in neutron-rich even-$N$ (odd-$N$) Au isotopes reveal the following features:
\begin{enumerate}[leftmargin=*]
\setlength\itemsep{0em}
\item[i)] The one-proton hole configuration $\pi 0h_{11/2}^{-1}$ dominates in both even-$N$ and odd-$N$ isotopes.
\item[ii)] In odd-$N$ isotopes, the configuration with an additional one-neutron hole $\nu 0i_{13/2}^{-1}$ is predominant.
\end{enumerate}
These results indicate that the wave functions of the observed Au states are characterized by the $\pi 0h_{11/2}^{-1}$ ($\pi0h_{11/2}^{-1}\nu0i_{13/2}^{-1}$) hole configuration built on the Hg core (see Fig.~\ref{fig:Occ}).

A question arises whether the similarities of the observed energies of Au isotopes from the corresponding yrast states in the Hg core follows an intuitive weak-coupling/decoupling scheme of the particle-core coupling model.
Without invoking any explicit assumptions about the collective motion of the Hg cores, we analyzed the LSSM wave functions of the Au isotopes in terms of hole(s) coupled to the states of the corresponding Hg core nucleus.
Explicitly, a state $|\text{Au}(J_{\rm oe})\rangle$ in an odd-even (oe) Au isotope is expanded in terms of the particle-core coupled components $|{\rm Hg}(J_{{\rm c},i}),j_\pi^{-1};J_{\rm oe}\rangle$, where $|J_{{\rm c},i}\rangle$ is a $i^{th}$ state with spin $J$ in Hg and $|j_\pi^{-1}\rangle$ is a hole in the proton orbital $j_\pi$, both of which are coupled to angular momentum $J_{\rm oe}$. 
This decomposition has the components
\begin{equation}
\alpha_{J_{{\rm c},i}}^{J_{\rm oe}}(j_\pi)=
c\langle{\rm Au}(J_{\rm oe})||
\tilde a_{j_\pi}||
{\rm Hg}(J_{{\rm c},i})\rangle
\end{equation}
where $\langle{\rm Au}(J_{\rm oe})||\tilde a_{j_\pi}||{\rm Hg}(J_{{\rm c},i})\rangle$ is a reduced matrix element obtained from the Wigner-Eckart theorem defined according to the convention of Ref.~\cite{Talmi13}.
The proportionality coefficient $c$ is obtained from the normalization condition
$\sum_{J_{{\rm c},i}j_\pi}\bigl(\alpha_{J_{{\rm c},i}}^{J_{\rm oe}}(j_\pi)\bigr)^2=1$,
where the sum is over all core states $J_{{\rm c},i}$
and proton orbitals $j_\pi$ included in the LSSM calculation.
The coefficient $\alpha_{J_{{\rm c},i}}^{J_{\rm oe}}(j_\pi)$ represents the overlap between the wave function of the Au and the corresponding Hg states, with $|\alpha_{J_{c,i}}^{J_{\rm oe}}|^2$ being related to the spectroscopic factor.
Similarly, high-spin states in odd-odd Au isotopes can be expanded on the basis that a neutron-proton hole pair couples to the Hg core states. 

\begin{figure}[t!]
	\centering 
	\includegraphics[width=\columnwidth]{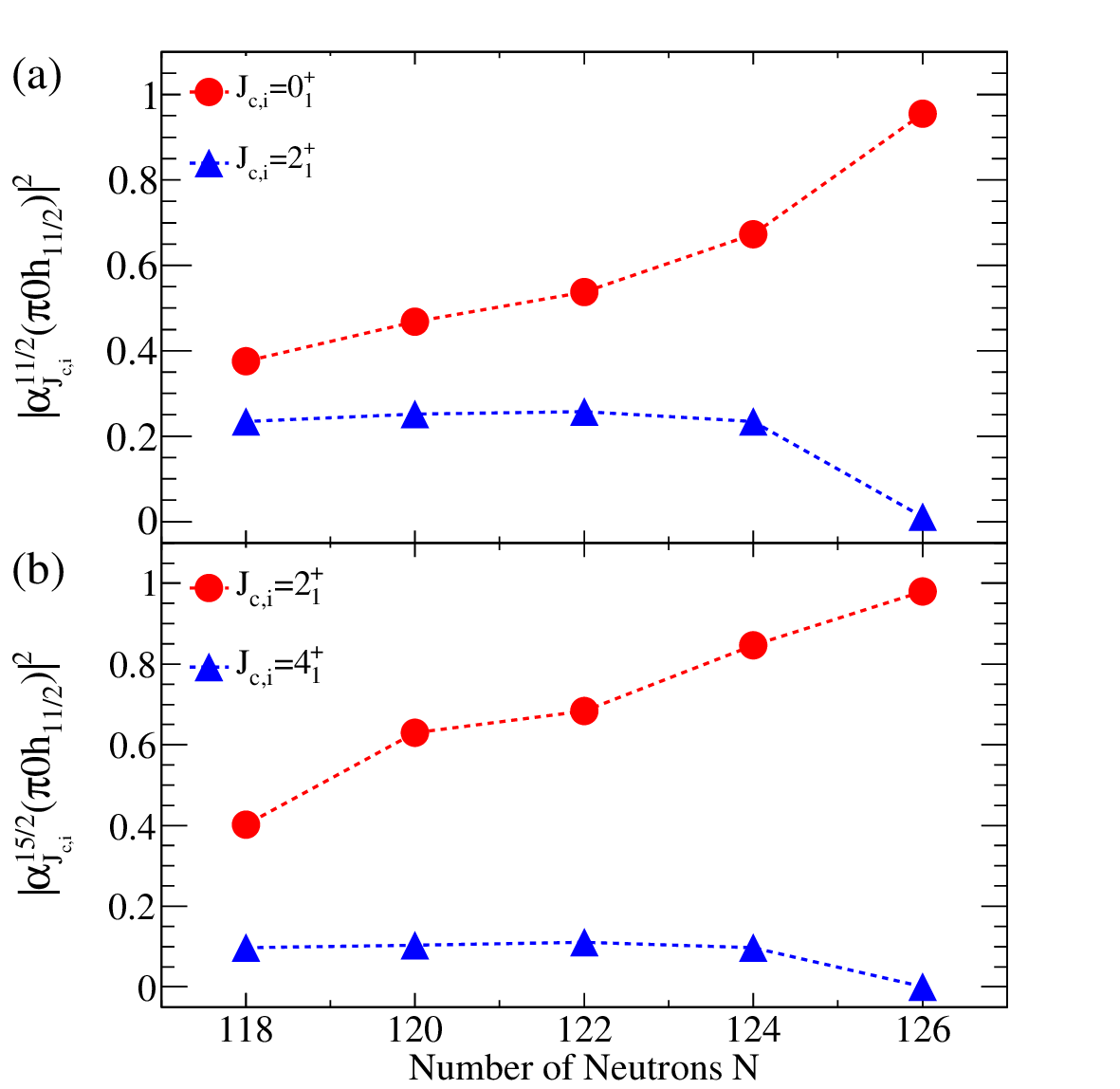}		
	\caption{
      The overlap $|\alpha_{J_{c,i}}^{J_{oe}}|^2$ as a function of neutron number $N$ for (a) $J_{\rm oe}^-=11/2_1^-$ in Au
     with $J_{c,i}=0_1^+$ (red circles) and $J_{c,i}=2_1^+$ (blue triangles) in Hg. (b) same as (a) for $J_{\rm oe}^-=15/2_1^-$ with $J_{c,i}=2_1^+$ and $J_{c,i}=4_1^+$ in Hg .
     }
    \label{fig:wf_overlap}
\end{figure}

Fig.~\ref{fig:wf_overlap} shows the overlaps for $11/2_1^-$ ($15/2_1^-$) in Au with the corresponding $0^+$ and $2^+$ ($2^+$ and $4^+$) states in Hg, respectively. 
These overlaps are given by the squares of $\alpha_{0_1^+}^{11/2^-}(\pi0h_{11/2})$ and $\alpha_{2_1^+}^{15/2^-}(\pi0h_{11/2})$, respectively, and are plotted for even-$N$ Au isotopes as a function of neutron number $N$.
As can be seen in Fig.~\ref{fig:wf_overlap}, for the $11/2_1^-$ state, $|\alpha_{0_1^+}^{11/2^-}(\pi0h_{11/2})|^2$ is close to 1 only near the  $N=126$ shell closure, indicating that overlap with the Hg $0_1^+$ state is dominant, consistent with a weak-coupling/decoupling picture. 
However, this intuitive structure rapidly breaks down as the neutron number decreases away from the $N=126$ shell closure. 
This can be seen from substantial admixtures from other-spin components in Hg core in a given state in Au.
As shown in Fig.~\ref{fig:wf_overlap}(a), the overlap with Hg($0_1^+$) decreases sharply, while the overlap with the higher-spin component Hg($2_1^+$) increases and becomes comparable.
Contributions from other excited states, such as Hg($0_2^+$), remain negligible.
A similar behavior is observed for the $15/2_1^-$ states (Fig.~\ref{fig:wf_overlap}(b)), although the mixing with the higher spin state is less pronounced than for the $11/2_1^-$ state.
These results indicate that even though the observed levels in the odd-even (odd-odd) Au isotopes arise from $\pi 0h_{11/2}^{-1}$ ($\pi0h_{11/2}^{-1}\nu0i_{13/2}^{-1}$) configurations coupled to the corresponding Hg core, they do not conform to a weak-coupling/decoupling picture. 
The above presented microscopic analysis based on the LSSM calculations demonstrates for the first time, that the observed regularity and inheritance highlight the role of large configuration mixing of the Hg core,  unlike what is expected from the conventional weak-coupling/decoupling interpretation in the Pb region.

To elucidate the origin of regular level-energy patterns in relation to $N$ observed in Au–Hg–Tl isotopic chains, restricted-space shell-model calculations were carried out for Hg isotopes. 
In these restricted calculations, the dominant proton configurations in Hg, $\pi d_{3/2}^4$ or $\pi d_{3/2}^2s_{1/2}^2$, were coupled separately to each neutron orbital ($0i_{13/2}$, $0h_{9/2}$, $1f_{7/2}$, or $1f_{5/2}$). The calculated excitation energies of the $0^+$, $2^+$, and $4^+$ states in Hg are nearly independent of neutron occupancies for each orbital with similar energy. 
This indicates that the level energy is almost insensitive to smooth changes in neutron occupancies and configurations.
The LSSM calculated number of holes in each orbital slowly decreases for all normal parity orbitals ($\nu 0h_{9/2}$, $\nu 1f_{5/2}$, $\nu 2f_{7/2}$, $\nu 2p_{1/2}$, and  $\nu 2p_{3/2}$) as a function of neutrons (Fig.~\ref{fig:Occ}), where the Hg isotopes show similar occupancy.
The insensibility of the level energies to the occupancy of the neutron orbitals accounts for the observed regularity of the level energies in the Hg and consequently in Au and Tl, which follows the Hg systematics. 

In conclusion, hard-to-reach neutron-rich isotopes of Au were populated through multi-nucleon transfer reactions at near Coulomb barrier energies. New high-spin states in $^{196-202}$Au were identified for the first time using the VAMOS++ large-acceptance spectrometer for projectile-like fragments, the CATLIFE detection system for target-like fragments, and the AGATA $\gamma$-ray tracking array for prompt $\gamma$ rays. The measured relative excitation energies of these high-spin states exhibit remarkable regularities and a clear inheritance pattern from the corresponding Hg cores.
Simplified shell-model calculations reproduce the observed regular level-energy pattern in both Hg and Au isotopes. 
Large-scale shell-model calculations further reveal that Au wave functions are dominated by $\pi 0h_{11/2}^{-1}$ ($\pi0h_{11/2}^{-1}\nu0i_{13/2}^{-1}$) configurations coupled to the Hg core. 
However, in contrast to an expected intuitive interpretation of a weak coupling/decoupling limit of the particle-core coupling model, Au configurations exhibit significant mixing with higher-spin components of the corresponding Hg core, as they move away from the $N=126$ shell closure.
Additionally, the intruder orbital-coupled bands in Au demonstrate that level inheritance may occur from the $(Z+1)$ or $(Z-1)$ core, depending on whether the core couples to valence or intruder orbital excitations. 
These findings provide new insights into the evolution of the shell structure and the configuration mixing in the vicinity of the $N=126$ shell closure.
Systematic investigations into the isotopes with inheritance from the neighboring core could shed more light on how the valence particles/holes are coupled to the core and help to better understand the underlying interaction driving the inheritance.

\section*{Acknowledgements}

The authors are grateful for the excellent support from the GANIL accelerator staff and the AGATA collaboration. 
YC, YS, SC acknowledge the support from the National Research Foundation of Korea (NRF) grants funded by the Korean government (MSIT), Grants No. RS-2022-NR070152 and RS-2024-00436392. 
YC, YHK, KIH, YS, JP, JH and SB acknowledge the support from the Institute for Basic Science (IBS) of the Republic of Korea (IBS-R031-D1). 
AA acknowledges that the work was supported by the UK STFC and GANIL visiting researcher program.
CHK, MJK, GMG, and KC were supported by the National Research Foundation of Korea (RS-2024-00338255). 
JH was supported by the IBS of the Republic of Korea (grant No. IBS-R031-Y3) and the National Research Foundation of Korea funded by the Ministry of Education (grant no. 2020R1A6A3A03039081). 
CY and ML acknowledge the National Natural Science Foundation of China under Grant No. 12475129.  
RPV acknowledges work partially funded by MCIN/AEI/10.13039/501100011033, Spain, with grants PID2020-118265GB-C4, PID2023-150056NB-C4, PRTR-C17.I01, by Generalitat Valenciana, Spain, with grant CIPROM/2022/54, ASFAE/2022/031, and by the EU NextGenerationEU and FEDER funds.

\bibliographystyle{elsarticle-num} 
\bibliography{Bib1}
\end{document}